\documentclass[12pt]{iopart}
\usepackage{graphicx}
\graphicspath{{}}
\begin{document}
	
\title[Intense, cold, velocity-controlled molecular beam by frequency-chirped laser slowing]{An intense, cold, velocity-controlled molecular beam by frequency-chirped laser slowing}
	
\author{S. Truppe, H. J. Williams, N. J. Fitch, M. Hambach, T. E. Wall, E. A. Hinds, B. E. Sauer and M. R. Tarbutt,}
\address{Centre for Cold Matter, Blackett Laboratory, Imperial College London, Prince Consort Road, London SW7 2AZ, UK}
\ead{m.tarbutt@imperial.ac.uk}

\begin{abstract}
Using frequency-chirped radiation pressure slowing, we precisely control the velocity of a pulsed CaF molecular beam down to a few m/s, compressing its velocity spread by a factor of 10 while retaining high intensity: at a velocity of 15~m/s the flux, measured 1.3~m from the source, is 7$\times$10$^{5}$ molecules per cm$^{2}$ per shot in a single rovibrational state. The beam is suitable for loading a magneto-optical trap or, when combined with transverse laser cooling, improving the precision of spectroscopic measurements that test fundamental physics.  We compare the frequency-chirped slowing method with the more commonly used frequency-broadened slowing method.
\end{abstract}

\maketitle

\section{Introduction}

Molecular beams with controllable forward velocity have been at the forefront of cold ($T\sim$~1--1000~mK) molecule research for many years~\cite{vandeMeerakker2012}.  Such beams are increasingly being used for precise measurements that test fundamental physics, including measurements of the electron's electric dipole moment~\cite{Hudson2011, Baron2014}, parity violation in nuclei~\cite{Cahn2014} and chiral molecules~\cite{Daussy1999, Tokunaga2013}, changes to the fundamental constants~\cite{Shelkovnikov2008, Hudson2006, Truppe2013} and tests of QED~\cite{Salumbides2011}. The precision of these measurements could be greatly improved using colder and slower molecular beams, preferably in the ultracold regime ($T\leq1$~mK).  Traditional techniques for controlling the forward velocity, such as Stark deceleration and its variants~\cite{Bethlem1999,Osterwalder2010,Fulton2004,Narevicius2008}, as well as recently-developed alternatives~\cite{Chervenkov2014,Fitch2016}, do not provide cooling. In some cases, molecules have been trapped and then cooled to lower temperatures by adiabatic~\cite{Perez2013}, evaporative~\cite{Stuhl2012} or Sisyphus~\cite{Zeppenfeld2009,Zeppenfeld2012,Prehn2016} cooling. Sympathetic cooling may also be possible~\cite{Tokunaga2011,Lim2015}.

Recently, a few molecular species have been directly laser cooled, either by compressing the transverse velocity distribution of a molecular beam~\cite{Shuman2010, Hummon2013}, or in a magneto-optical trap (MOT) which provides simultaneous trapping and cooling~\cite{Barry2014, McCarron2015, Norrgard2016}. An important current challenge is to increase the number of molecules in the MOT by increasing the fraction delivered below the capture velocity, which is typically 10--20~m/s \cite{Tarbutt2015}. At present, radiation pressure slowing is used~\cite{Barry2012}, with the laser linewidth broadened to address a wide velocity range~\cite{Barry2012,Yeo2015,Hemmerling2016}. This approach yields limited control of the final velocity and typically slows the beam without compressing the velocity distribution, delivering only a tiny fraction of the molecules at the desired position and speed.  Here, we present an alternative approach, using frequency-chirped laser slowing of CaF to both compress the velocity distribution into a narrow range and slow to the desired final velocity.  We find this approach superior to the frequency-broadened technique, realizing finer velocity control, decreased temperature, and greatly increased molecular flux, all of which are essential for making dense molecular MOTs and intense molecular beams for precise measurements.

\section{Experiment Setup}

\begin{figure}[tb]
	\centering
\includegraphics[width=0.75\columnwidth]{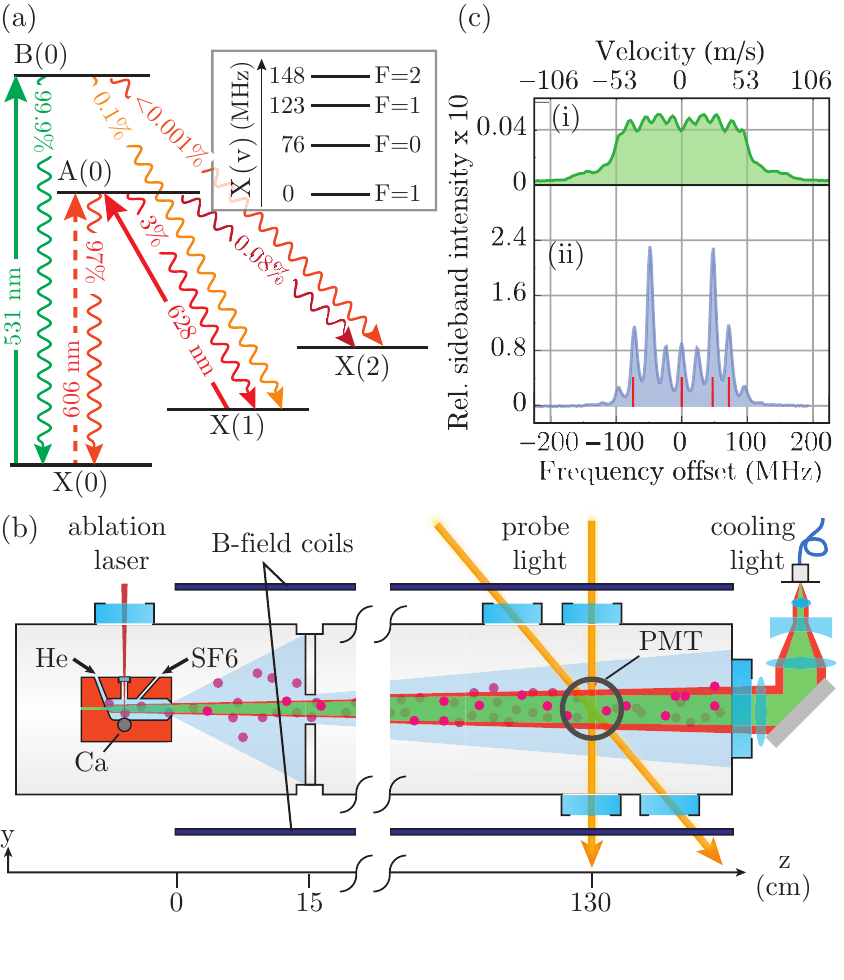}
\caption{\label{Setup} Relevant energy levels of CaF, with calculated vibrational branching ratios~\cite{Pelegrini2005, Dulick1980}, and the transitions used for slowing (solid lines) and detection (dashed line). Wavy lines are spontaneous decays. $v$, $N$, $J$, and $F$ are the vibrational, rotational, total electronic and total angular momentum quantum numbers, respectively. We use X$(v)$, A$(v)$ and B$(v)$ to denote the states X$^{2}\Sigma^{+}$($v$,$N$=1), A$^{2}\Pi_{1/2}$($v$,$J$=1/2,$p$=+1) and B$^{2}\Sigma^{+}$($v$,$N$=0) respectively, where $p$ is the parity. Inset: hyperfine structure in X(0)~\cite{Childs1981}. The hyperfine interval of B(0) is 20(5)~MHz and of A(0) is $<$10~MHz~\cite{Tarbutt2015}. 
(b) Apparatus. A pulsed cryogenic beam of CaF is slowed by a counter-propagating laser beam.  Molecules are detected by LIF at $z$=130~cm, using a probe laser at either 90$^{\circ}$ or 60$^{\circ}$ to the molecular beam. The probe lasers have gaussian intensity distributions with $1/e^2$ diameters of 6~mm. (c) Spectrum of main cooling laser with  frequency-broadened (i) and frequency-chirped (ii) light, measured by a spectrum analyzer with 10~MHz linewidth. Intensities are relative to the unmodulated light. Lines mark frequencies of hyperfine components.}
\end{figure}

Figure~\ref{Setup}(a) shows the relevant energy levels of CaF and the vibrational branching ratios between them, along with our notation. 
The main cooling transition is B(0)--X(0) with wavelength $\lambda_{\rm{main}}$=531~nm, linewidth $\Gamma$=2$\pi\times$6.3~MHz~\cite{Dagdigian1974} and single-photon recoil velocity 1.3~cm/s. Population that leaks into X(1) is returned to the cooling cycle via the A(0)--X(1) transition at $\lambda_{\rm{repump}}$=628~nm. From an experimental study of potential loss channels (see Sec.~\ref{losses}), we conclude that with only these two wavelengths, $\sim$3$\times$10$^{4}$ photons per molecule can be scattered, corresponding to a velocity change of 390~m/s, before half are lost from the cooling cycle.  Using separate upper states for the main cooling and repump lasers almost doubles the scattering rate \cite{Tarbutt2013b} relative to all previous work~\cite{Shuman2009, Shuman2010, Barry2012, Hummon2013, Zhelyazkova2014, Yeo2015, Hemmerling2016} where X(0) and X(1) were both driven to A(0).

Figure~\ref{Setup}(b) illustrates the apparatus. A pulsed beam of CaF is produced by a cryogenic buffer gas source~\cite{Hutzler2012, Barry2011a, Bulleid2013}. At $t$=0, a pulsed laser (5~mJ, 4~ns, 1064~nm) ablates Ca into a 4~K copper cell, through which flow 1~sccm of 4~K helium and 0.01~sccm of 270~K SF$_6$. The resulting CaF molecules are cooled by the He and entrained in the flow. They exit the cell at $z$=0 via a 3.5~mm diameter aperture, and are collimated by an 8~mm diameter aperture at $z$=15~cm that separates the source from the main chamber, where the pressure is 3$\times$10$^{-7}$~mbar. Within a factor of 2, the flux is 1.9$\times$10$^{11}$ molecules per steradian per shot in X(0), and the pulse duration at $z$=2.5~cm is 280~$\mu$s (FWHM). At $z$=130~cm the molecules are detected by driving the A(0)$\leftarrow$X(0) transition, imaging the resulting laser-induced fluorescence (LIF) onto a photomultiplier tube (PMT), and recording the signal with a time resolution of 5~$\mu$s, yielding a time-of-flight (ToF) profile. The 5~mW probe beam crosses the molecular beam at 60$^{\circ}$ or 90$^{\circ}$ to the molecular beam propagation direction for velocity-sensitive or insensitive measurements, respectively.
Radio frequency sidebands applied to the probe~\cite{Zhelyazkova2014} address the four hyperfine components of the transition.

The cooling light counter-propagates to the molecular beam and consists of 110~mW at $\lambda_{\rm{main}}$ applied for times between $t_{\rm{start}}$ and $t_{\rm{end}}$, and 100~mW at $\lambda_{\rm{repump}}$, which is applied continuously. The two wavelength components have orthogonal linear polarizations, both at 45$^\circ$ to a uniform 0.5~mT magnetic field directed along $y$, which prevents optical pumping into dark Zeeman sub-levels~\cite{Berkeland2002, Stuhl2008,Shuman2009}. For most experiments, the cooling light is collimated and has a gaussian intensity distributions with $1/e^2$ diameter of 6~mm. For the experiments described in Sec.~\ref{superSlow}, the light converges towards the molecular source. The main cooling light is blocked on alternate experimental shots so that measurements with and without cooling can be compared. To address all hyperfine components, we generate the spectrum shown in Fig.~\ref{Setup}(c,ii) by passing both lasers through electro-optic modulators (EOMs) driven at 24~MHz with a modulation index of 3.1.
We find the frequencies, $f_{\rm{main}}$ and $f_{\rm{repump}}$, that maximize the LIF when each laser in turn is used as an orthogonal probe. Then we detune the two cooling lasers so that, when counter-propagating to the molecules, they are resonant with those travelling with speed $v_{\rm{start}}$. To compensate the changing Doppler shift as the molecules slow down, we apply linear frequency chirps with rates $\beta$ and $\beta \lambda_{\rm{main}}/\lambda_{\rm{repump}}$ to the main and repump lasers, respectively.  
To compare this frequency-chirped method with the frequency-broadened method used in previous work~\cite{Barry2012,Yeo2015,Hemmerling2016}, we fix the centre frequencies at $f_{\rm{main}} - f_{\rm{offset}}$ and $f_{\rm{repump}} - f_{\rm{offset}}\lambda_{\rm{main}}/\lambda_{\rm{repump}}$, and produce the broadened spectrum shown in Fig.~\ref{Setup}(c,i) by sending the light through three consecutive EOMs driven at 72, 24, and 8~MHz.

\section{Method for determining velocity distributions}
\label{analysisSection}

To determine a velocity distribution, we compare the Doppler-shifted spectrum recorded using the 60$^{\circ}$ probe laser with the unshifted spectrum recorded using the 90$^\circ$ probe. In principle, the velocity distribution could be extracted directly from a comparison of these spectra. There are three disadvantages to this direct method. First, the spectrum has hyperfine structure that spans roughly the same frequency interval as the Doppler shifts, and this complicates the conversion of the spectrum into a velocity distribution. Second, the spectral resolution limits the velocity resolution to about 20~m/s. While this can be improved upon by deconvolving the spectral profile recorded using the 90$^\circ$ probe, that introduces additional noise. Third, the method does not make use of all the available information, in particular the fact that there is a strong correspondence between velocity and arrival time. Instead, we employ a novel analysis method where we first determine that correspondence, and then use it to convert the ToF profile to a velocity distribution.

\begin{figure}[tb]
	\centering
	\includegraphics[width=0.8\columnwidth]{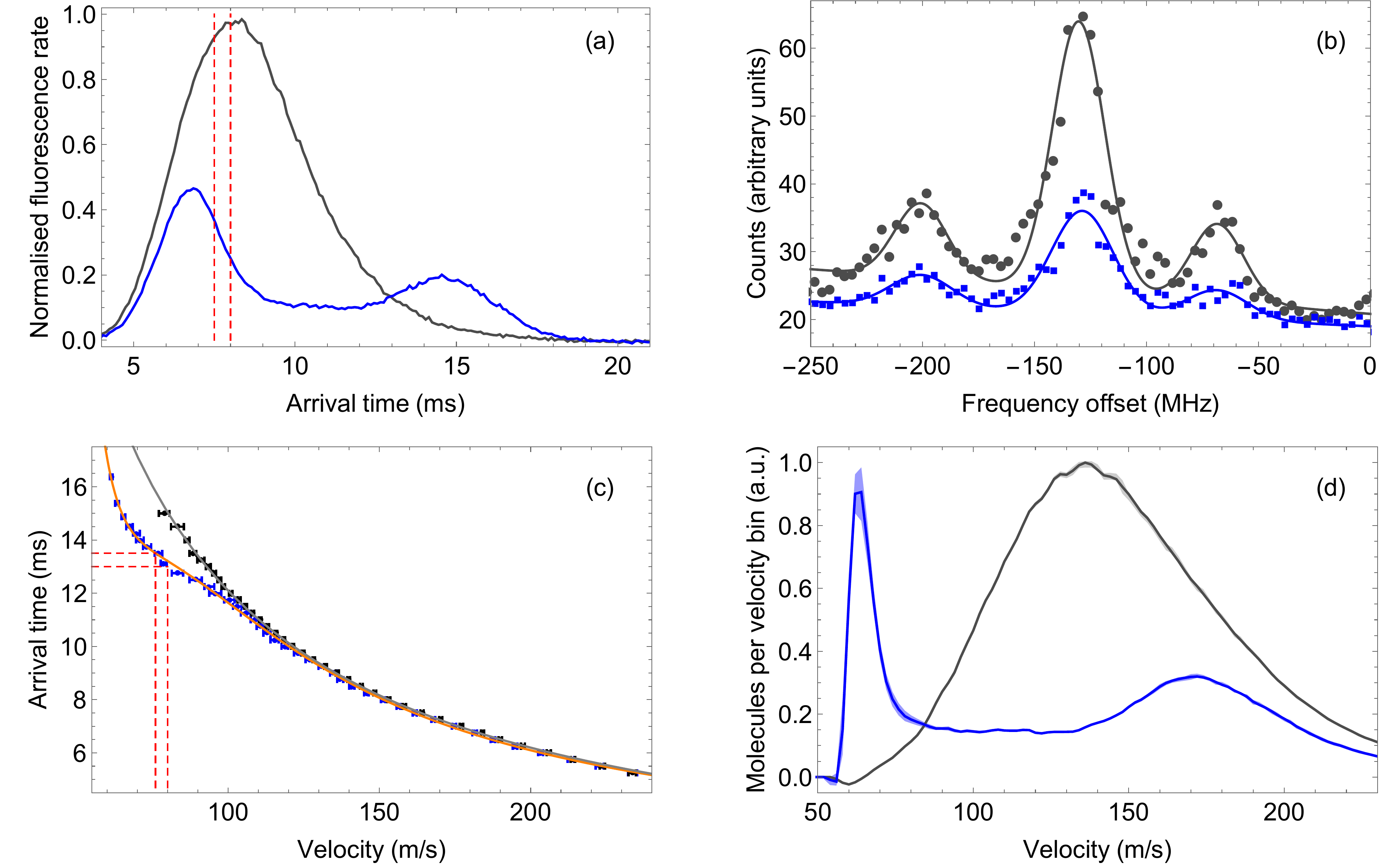}
	\caption{\label{AnalysisMethod} Method for determining the velocity distribution, illustrated for data with $\beta = 21$\,MHz/ms, $t_{\rm{start}}=1$\,ms, $t_{\rm{end}}=7$\,ms, $v_{\rm{start}}=178$\,m/s. Throughout, blue and grey data have cooling light on and off respectively. (a) Control and cooled ToF profiles recorded using the 90$^\circ$ probe. (b) Spectrum recorded using the 60$^\circ$ probe, for molecules arriving in the 7.5--8\,ms time window [the region between the dashed lines in (a)]. The Doppler shift determines the mean velocity of molecules arriving in this time window. Dots: data. Lines: fit to sum of three gaussians. (c) Dots: arrival time versus mean velocity determined this way. The error bars are obtained from the fit to the spectrum. Lines: fits to the model described in the text. The number of molecules in a velocity bin, such as the one between the dashed lines, is found by reading off the corresponding time bin and then integrating the ToF profile within that time bin. (d) Velocity distributions obtained by this method. The coloured bands around the solid lines indicate the 68\% confidence limits determined using the method described in the text.}
\end{figure}

Figure~\ref{AnalysisMethod} illustrates the analysis method using data with $\beta = 21$\,MHz/ms, $t_{\rm{start}}=1$\,ms, $t_{\rm{end}}=7$\,ms, and $v_{\rm{start}}=178$\,m/s. Data with the cooling light off (on) is referred to as ``control'' (``cooled''). Figure~\ref{AnalysisMethod}(a) shows the control and cooled ToF profiles recorded using the 90$^\circ$ probe, each averaged over 50 shots. To measure the velocity profile we first record a Doppler-free reference spectrum using the 90$^{\circ}$ probe. The peak fluorescence signal in this spectrum defines the zero of frequency. We then measure a velocity-sensitive spectrum using the 60$^{\circ}$ probe. We partition this data by arrival time, using 0.5\,ms-wide time windows, so that the range of velocities is small and the spectrum is similar to the reference spectrum, but shifted according to the mean velocity. Figure~\ref{AnalysisMethod}(b) shows the control and cooled spectra for molecules arriving between 7.5 and 8\,ms, the time window indicated by the dashed lines in (a). Because there are four hyperfine components, and the light has four rf sidebands, there are several peaks in the spectrum, three of which are clear in the data. The largest peak is obtained when the four hyperfine components are simultaneously resonant. We fit the data to a sum of three gaussians and use the fitted centre frequency of the largest peak to determine the mean velocity. The uncertainty in this mean velocity is also obtained from this fit. Applying this procedure to all time windows gives graphs of arrival time versus mean velocity, as in Fig.~\ref{AnalysisMethod}(c). We use these measured correlations between velocity and arrival time to turn the ToF profiles into velocity distributions. To do that we need to join the points, and we have experimented with three different ways of doing this, all of which produce very similar velocity distributions. The simplest is linear interpolation. This works well but is not ideal because the gradient is discontinuous at each data point and the conversion between distributions is proportional to this gradient. It is preferable to represent the data by a smooth curve, and we find that construction of a B-spline function can achieve that and also works well. The third method, and the one we favour, is to fit the model $t = \sum_{n=0}^{m}{a_n/v^n}$ to the data, where $a_{n}$ are free parameters and we choose $m$ appropriately. We choose to use this method for all our data, since it works well and allows us to use standard fitting algorithms and goodness-of-fit measures. The control data fits well with $m=1$, as expected for zero deceleration. For the cooled data, we take $m=5$ since this gives an adequate fit for all the datasets. For the data in Fig.~\ref{AnalysisMethod}(c) this is the smallest value of $m$ where $\chi^{2}$ is smaller than the median of the chi-squared distribution. To find the number of molecules with velocities in the range $v\pm\Delta v$, we use the curves of Fig.~\ref{AnalysisMethod}(c) to find the times $t_{1,2}$, corresponding to $v\pm\Delta v$ with $\Delta v=2$\,m/s, then integrate the ToF profile between $t_{1}$ and $t_{2}$. Doing this for all velocities gives the control and cooled velocity distributions such as those shown in Fig.~\ref{AnalysisMethod}(d). 

To determine a statistical confidence interval, we proceed as follows. For each data point in Fig.~\ref{AnalysisMethod}(c) we generate 400 new velocity values drawn at random from a normal distribution with mean and standard deviation given by the central value and error of that data point. From these, we construct 400 new time-versus-velocity curves and associated velocity distributions using exactly the same method as described above. From this large set of velocity profiles, we find the mean value at each point, along with the upper and lower limits that bound 68\% of the values above and below the mean. Finally, all the profiles are divided by the maximum value of the control profile, so that the peak of every control profile is set to 1. The solid lines in Fig.~\ref{AnalysisMethod}(d) show the mean profiles, and the bands around them represent the 68\% confidence interval. The accuracy of our analysis method is discussed in detail in \ref{app}.

\section{Results}

\subsection{Frequency-chirped slowing}

\begin{figure}[tb]
	\centering
\includegraphics[width=0.9\columnwidth]{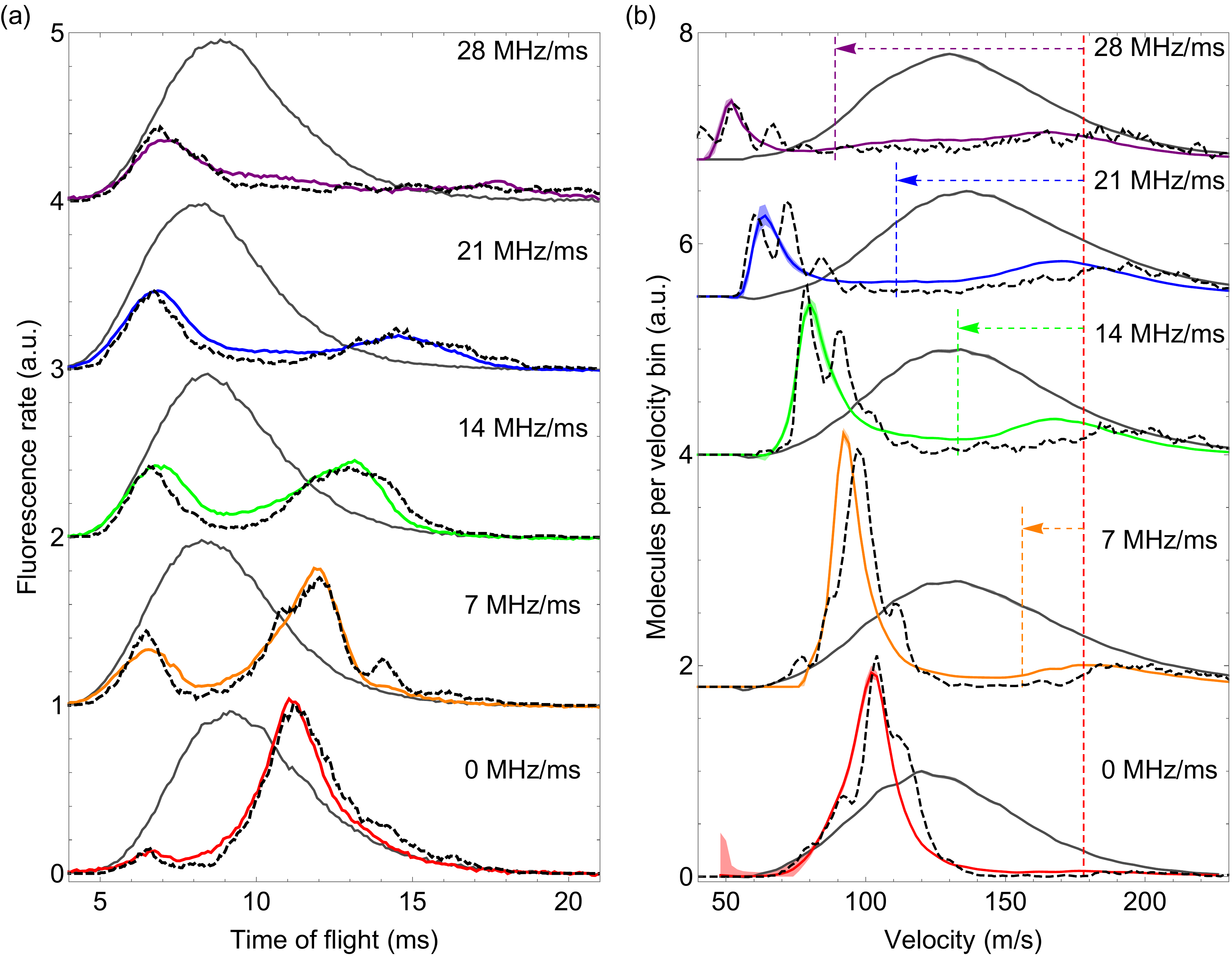}
\caption{\label{Chirp_Data} Laser slowing for various chirp rates, $\beta$. Other parameters are $t_{\rm{start}}$=1~ms, $t_{\rm{end}}$=7~ms, $v_{\rm{start}}$=178~m/s. (a) ToF profiles and (b) velocity distrbutions, with cooling off (solid grey) and on (solid coloured). Black dashed curves are simulation results. Vertical dashed lines in (b) show $v_{\rm{start}}$ (red) and $v_{\rm{end}}$ (other colours). Coloured bands around the solid lines in (b) indicate 68\% confidence limits.}
\end{figure}

The solid curves in Fig.~\ref{Chirp_Data} are experimental control and cooled ToF profiles and velocity distributions for various chirp rates, with $t_{\rm{start}}$=1~ms, $t_{\rm{end}}$=7~ms, and $v_{\rm{start}}$=178~m/s. When $\beta$=0, the molecules are slowed to about 100~m/s and their velocity distribution is compressed. This is reflected in the ToF profile as a depletion at early times and an enhancement at later times. As $\beta$ increases, the molecules are pushed to lower velocities, and while they arrive at the detector over a broad range of times, they always have a narrow velocity distribution. The widths of the slow peaks correspond to a temperature of about 100~mK. The final velocity is always lower than $v_{\rm{end}}$, indicating that the molecules follow the changing frequency up to the highest $\beta$ used. The dashed curves in Fig.~\ref{Chirp_Data} are simulation results. For each simulation, we use a rate model~\cite{Tarbutt2015b} to determine the scattering rate versus detuning and power, and then calculate the resulting trajectories of many molecules using the experimental parameters and measured initial velocity distributions as inputs. The randomness of the momentum kicks is included. 

For all $\beta$, the simulations accurately predict the observed ToF profiles and velocity distributions, including the overall loss of detected molecules (see below). Some predicted structure in the slowed peak is not observed experimentally, but all other features agree well, showing that the scattering rate is as expected and the experiment is well understood. Supplementary simulations of a ten times longer molecular pulse, typical of most buffer-gas sources~\cite{Hutzler2012, Barry2011a}, indicate there is no difference in the velocity distribution or the tail of the ToF profile where the slow molecules arrive, provided the light is turned on once the majority of molecules have left the source. This shows that similar slowing performance can be expected for sources with more typical properties.

We find that the slowing depends critically on the applied magnetic field that remixes dark states. In the absence of this field the slowing light has no effect.  The deceleration increases with applied field up to 0.5~mT, corresponding to an average Zeeman shift of 3~MHz, where the effect saturates. Switching the polarization of the light~\cite{Berkeland2002, Hummon2013} at 5~MHz, with no applied magnetic field, gives the same results as a static polarization and a 0.5~mT magnetic field. Increasing the laser intensity increases the deceleration and the number of molecules decelerated, until the intensity reaches $\approx$350~mW/cm$^2$ where the effect saturates. 

\subsection{Frequency-broadened slowing}

\begin{figure}[tb]
	\centering
\includegraphics[width=0.9\columnwidth]{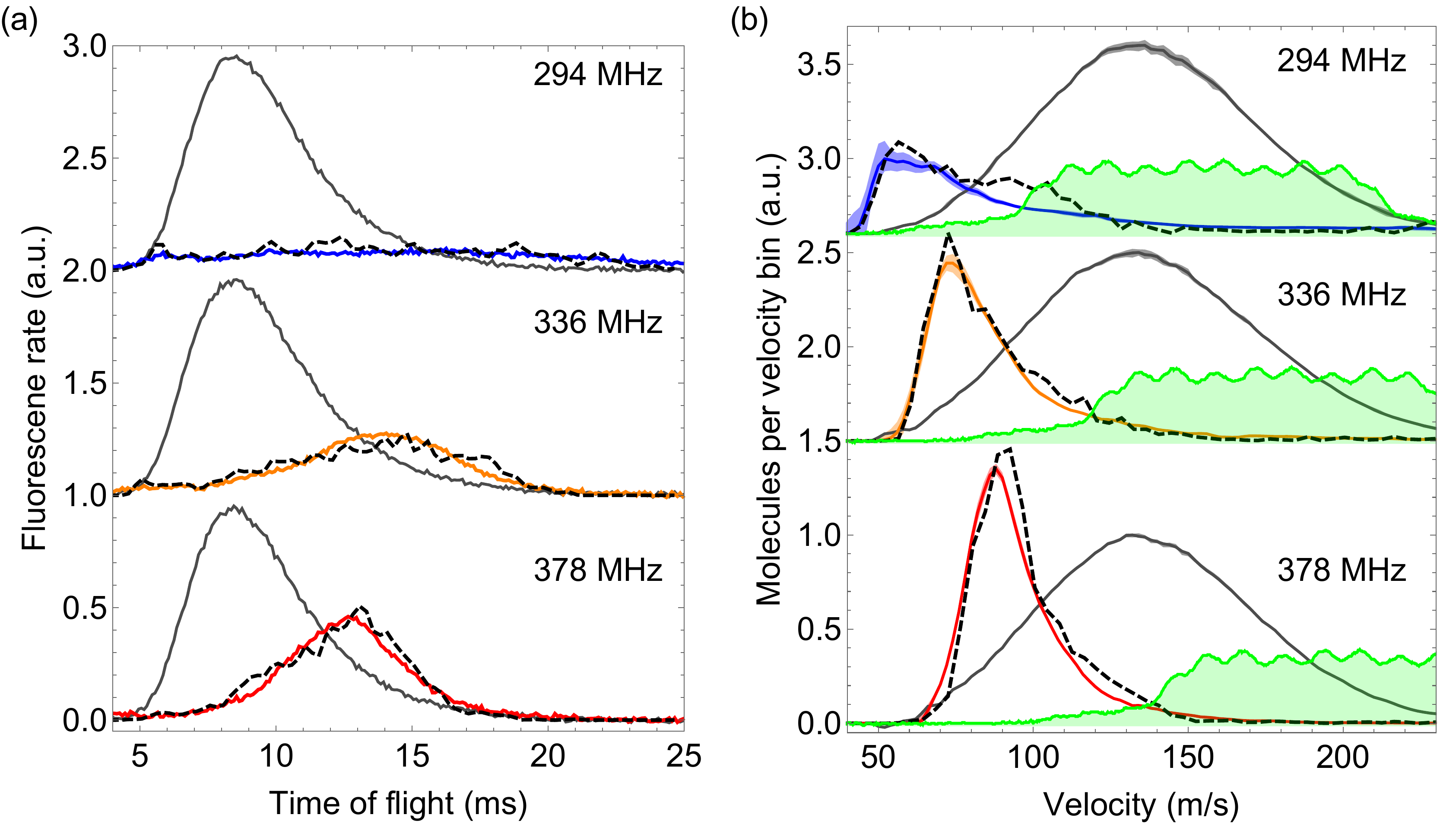}
\caption{\label{whiteLight} Laser slowing using frequency-broadened light with various values of $f_{\rm{offset}}$. Parameters are $t_{\rm{start}}$=1~ms, $t_{\rm{end}}$=7~ms, $\beta$=0. (a) ToF profiles and (b) velocity distrbutions, with cooling off (solid grey) and on (solid coloured). Black dashed curves are simulation results. Shaded area in (b): spectrum of light used in each case. Coloured bands around the solid lines in (b) indicate 68\% confidence limits.}
\end{figure}

For comparison with our frequency-chirped results, Fig.~\ref{whiteLight} shows ToF profiles and velocity distributions obtained using frequency-broadened light for three values of $f_{\rm{offset}}$. Again, we address most of the molecules and slow them efficiently. The velocity distribution is not as narrow as in the chirped case, but it is compressed. Though not seen in previous work, this is expected~\cite{Zhu1991} because all molecules are slowed until their Doppler shift is slightly below the low frequency cut-off of the broadened laser spectrum.  The simulations (dashed lines) agree very well with the measured ToF and velocity distributions, showing that this case is also well understood.

Just as for the chirped case, for the slowing to work it is essential to apply a magnetic field or to modulate the polarization of the light. Once again, we found that the deceleration increases with applied field up to 0.5~mT, and that switching the polarization of the light at 5~MHz has the same effect as a 0.5~mT magnetic field. The slowing saturates at a laser intensity of $\approx$750~mW/cm$^2$, about double the intensity needed for the chirped method.

\subsection{Losses}\label{losses}

Both slowing techniques show a decrease in the number of detected molecules as the velocity is reduced. To understand the reason, we first investigate the loss channels that might take population out of the cooling cycle. The laser slowing experiments themselves provide a very sensitive way to do this. To determine the fraction that leaks to state $q$, we scan the probe laser over a transition from $q$ and measure the increase in fluorescence when the cooling light is applied. Here, we use all the same parameters as in the $\beta = 21$\,MHz/ms data shown in Fig.~\ref{Chirp_Data}. We determine the fraction $f(q)=\Delta P(q)/P_{0}$ where $P_{0}$ is the initial population in X(0) and $\Delta P(q)$ is the change in the population of $q$ induced by the slowing lasers. Using the A(2) $\leftarrow$ X(2) transition we find $f(v=2) = 3.7(1)\%$. The simulations reproduce this result when the B(0)--X(2) branching ratio is $1.5(3)\times 10^{-5}$. Using the Q(0) and Q(2) lines of the $A^2\Pi_{1/2}(v=0)\leftarrow X^2\Sigma^+(v=0)$ transition, we find $f(N=0) = 1.6(2)\%$ and $f(N=2) = 0.4(2)\%$, corresponding to branching ratios of $7(1)\times 10^{-6}$ to $N=0$ and $1.6(3)\times 10^{-6}$ to $N=2$. The most obvious route to these even-parity states is the decay chain B--A--X, though there are other possibilities, including magnetic dipole transitions which are sometimes surprisingly intense for molecules~\cite{Kirste2012}. With similar sensitivity, we searched for possible loss to $N=3$ induced by a term in the hyperfine Hamiltonian that couples states with $\Delta N = 2$, but found nothing. From all these measurements we conclude that $\sim 3\times 10^{4}$ photons per molecule can be scattered before half are lost from the cooling cycle, and that very little of the loss observed in Figs.~\ref{Chirp_Data} and \ref{whiteLight} is due to leaks out of the cooling cycle.

Instead, the loss is due to the increased divergence of the slower molecules, compounded by stochastic transverse heating, as observed previously~\cite{Barry2012}. This increased divergence reduces the fraction of slow molecules that pass through the detection volume. The excellent agreement between experiment and simulation confirms this, since there are no other loss mechanisms in the simulations. Repeating the simulation for $\beta$=21~MHz/ms with transverse heating turned off, we find that the transverse heating is responsible for only 8\% of the total loss. Therefore, the dominant loss mechanism is the natural increase in divergence when the molecules are slowed down without any change to their transverse velocity distribution.

\subsection{Slowing to velocities below the capture velocity of a MOT}
\label{superSlow}

\begin{figure}[tb]
	\centering
\includegraphics[width=0.9\columnwidth]{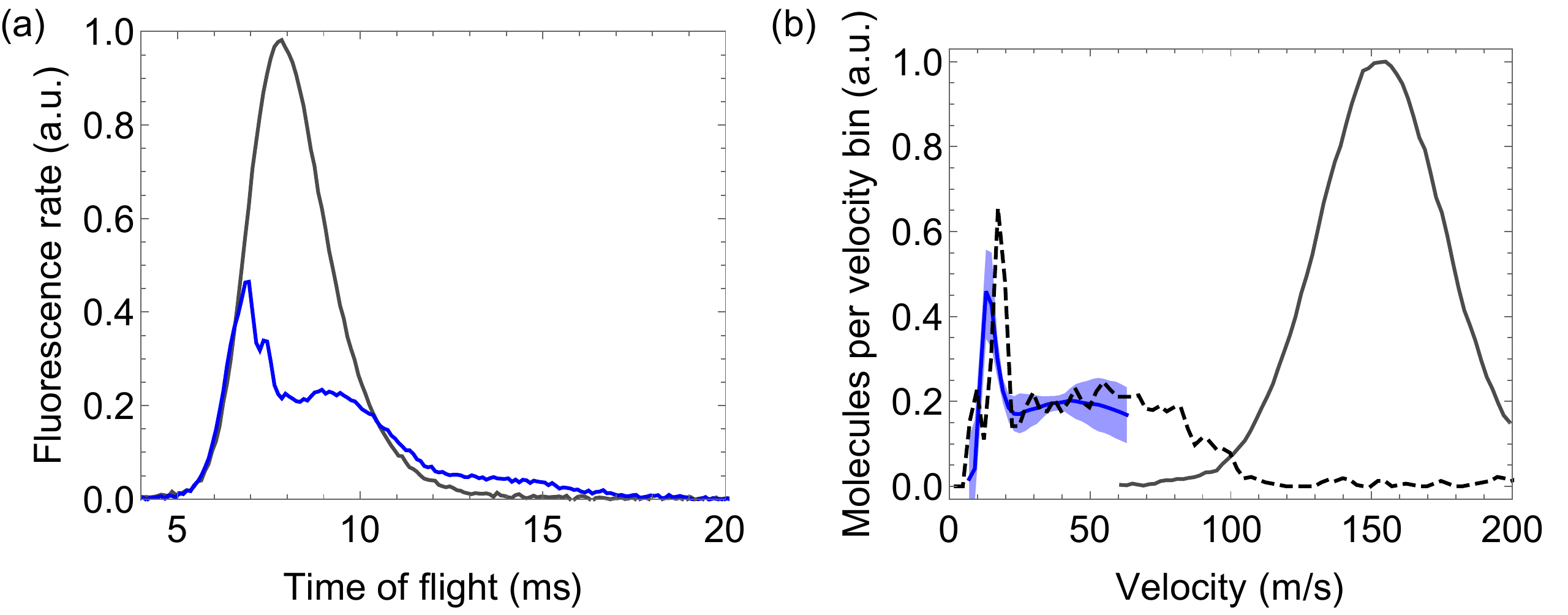}
\caption{\label{slowPeak} (a) ToF profile and (b) velocity distribution for cooling off (grey) and on (blue), with parameters $t_{\rm{start}}$= 3.5~ms, $t_{\rm{chirp}}$=4~ms, $t_{\rm{end}}$=12~ms, $\beta$=37~MHz/ms, $v_{\rm{start}}$=178~m/s, and a converging laser beam. Black dashed curves are simulation results. The coloured band in (b) indicates the 68\% confidence limits.}
\end{figure}

With the loss mechanisms understood, we increase the number of slow molecules in three ways. First, we add a small transverse force by converging the cooling beam with a full angle of 8.2~mrad to a $1/e^2$ diameter of 3~mm at $z$=0. This increases the number of detected molecules by 60\% relative to a collimated beam of the same power, using the same parameters as in Fig.~\ref{Chirp_Data} and $\beta$= 21~MHz/ms. Second, we reduce the free flight time for slowed molecules by increasing $t_{\rm{end}}$. 
Third, we change the chirp ramp so that the frequency is constant between $t_{\rm{start}}$ and $t_{\rm{chirp}}$, then linearly chirped between $t_{\rm{chirp}}$ and $t_{\rm{end}}$. This slows molecules with speeds greater than $v_{\rm{start}}$ before the chirp begins, so that they are no longer left behind, and increases the number of detected slow molecules by about 50\% when $t_{\rm{chirp}} - t_{\rm{start}}$=1~ms. Figure~\ref{slowPeak} shows the ToF profile and velocity distribution measured with these improvements. Molecules arriving between 12--16~ms all have mean speeds in the narrow range 15$\pm$2.5~m/s.  Within this range, the absolute number of molecules is 1$\times$10$^{6}$, the flux is 7$\times$10$^5$ molecules per cm$^{2}$ per shot, the intensity is 2$\times$10$^8$~cm$^{-2}$s$^{-1}$ and the brightness is 5$\times$10$^{9}$~cm$^{-2}$s$^{-1}$sr$^{-1}$, all to within a factor of 2. The velocity of these molecules is below the expected capture velocity of a MOT with $1/e^{2}$ beam diameters of 24~mm and readily available powers~\cite{Tarbutt2015}, indicating that $\approx$10$^{6}$ molecules per pulse could be loaded into a MOT.  The corresponding simulation agrees well with the data, being just 4~m/s faster and containing about 50\% more molecules.

\begin{figure}[tb]
	\centering
\includegraphics[width=0.7\columnwidth]{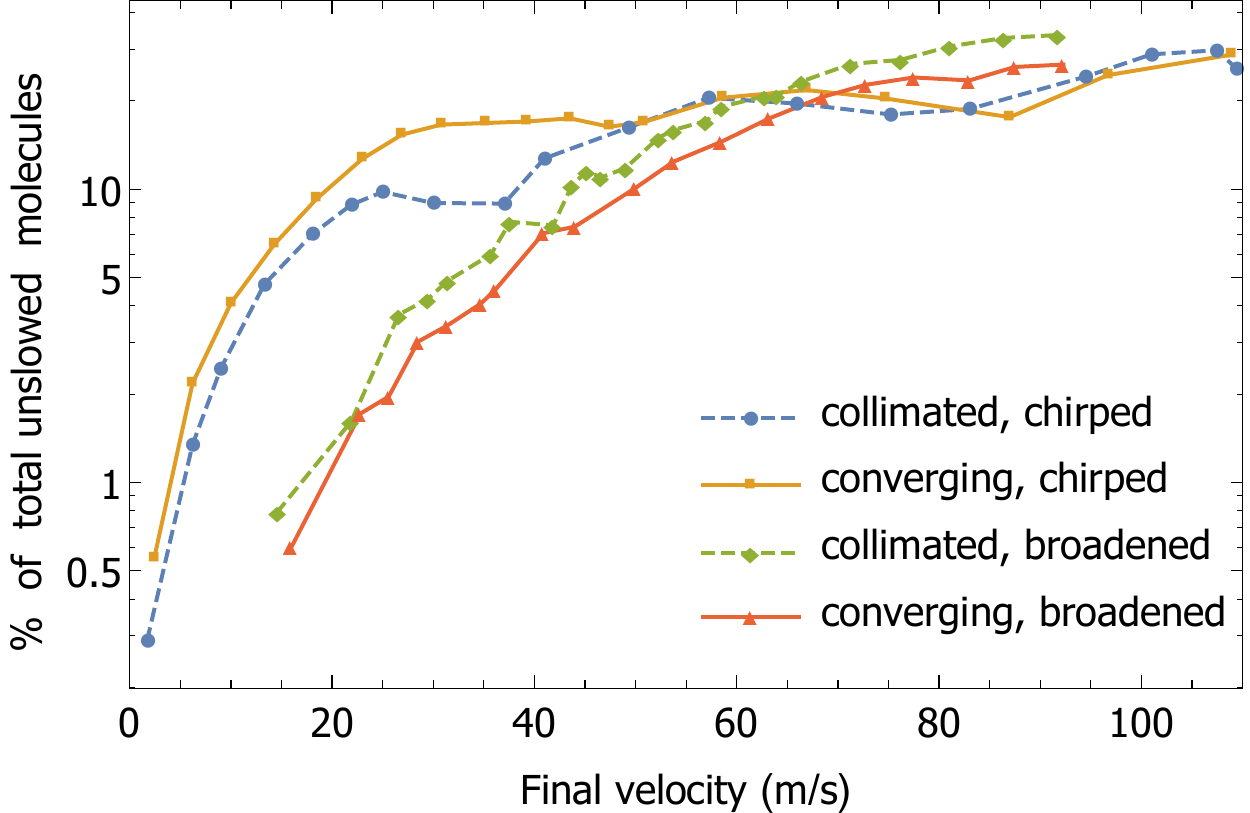}
\caption{\label{NumberVsVf} Comparing slowing methods: simulated number of slow molecules at the detector in a 10~m/s-wide interval centred on the peak velocity, as a function of that velocity. The number of slow molecules is expressed as a percentage of the total number of detected molecules in the control distribution. The velocity is controlled via $\beta$ (chirped case) and $f_{\rm{offset}}$ (broadened case), with the slowing light on between $t_{\rm{start}}$=4~ms and $t_{\rm{end}}=$12~ms.  All other parameters are the same as those for Figs.~\ref{Chirp_Data} and \ref{whiteLight}.}
\end{figure}

\subsection{Comparing the two slowing methods}

Figure~\ref{NumberVsVf} summarizes information from simulations where $\beta$ and $f_{\rm{offset}}$ are varied for the frequency-chirped and broadened cases respectively. We count the number of slow molecules at the detector in a 10~m/s-wide interval centred on the peak velocity, and plot this number versus that velocity. There is little difference between the two methods at higher velocities, but below 50~m/s the chirp method gives more slow molecules, e.g. about ten times more at 20~m/s. With broadened light, all molecules start slowing as soon as the light is turned on, those with high initial speeds never reach the final velocity, while those with low initial speeds reach it too early and then have a long way to travel with high divergence. For very low final speeds, these molecules may even come to rest before reaching the detector. The chirp method is more efficient because the slower molecules join the slowing process later on, and so a larger fraction of the initial distribution reaches the final velocity at a point close to the detector. Figure~\ref{NumberVsVf} also compares the effectiveness of the converging and collimated slowing beams. For frequency-broadened light converging the beam {\it reduces} the molecule number. This is because the slowing force has a low-velocity cut-off that shifts to higher velocities as $z$ increases, due to the falling light intensity, resulting in a much wider final velocity distribution: those that reach the cut-off early on have lower velocities than those that reach it later. Thus, while there are more molecules overall, there are fewer per unit velocity range. This does not happen in the chirped case, and so the converging beam yields an increase. 

\section{Optimization of slowing methods for MOT loading}

For the comparison shown in  Fig.~\ref{NumberVsVf}, the slowing light turn-on and turn-off times were chosen to be $t_{\rm{start}}$=4~ms and $t_{\rm{end}}$=12~ms, respectively.  While useful for comparing the various methods, this choice of parameters is generally not optimum for either of the slowing techniques. In simulations of frequency-broadened slowing, molecules reach their final velocity within 3--4~ms of the slowing light turning on.  After reaching a low enough velocity to fall out of resonance with the slowing light the molecules freely propagate to the detector at the slow final velocity and hence with a large divergence.  In contrast, when using frequency-chirped slowing, the forward velocity of the molecules tracks that of the chirp, decreasing linearly until the chirp ends.  In this case, molecules reach the final velocity at 12~ms and hence diverge less before reaching the detector.

A complete numerical optimisation of the laser power, convergence, turn-on time, turn-off time, initial frequency offset, and chirp rate (in the frequency-chirped case) involves too large a parameter space to be practical.  Instead, we fix the laser power and turn-off times at 100~mW and $t_{\rm{end}}$=12~ms, and vary the turn-on time $t_{\rm{start}}$. The beam convergence is fixed to one of two values, either ``collimated'' or ``converging''.  We also vary the offset frequency $f_{\rm{offset}}$ for frequency-broadened slowing, and  the chirp rate $\beta$ for frequency-chirped slowing. The initial frequency offset in the latter case is fixed at 335~MHz ($v_{\rm{start}}$=178~m/s).  For a metric to compare the simulation results over this limited parameter space, we choose the number of molecules that arrive at the MOT location with forward velocities below the expected capture velocity of $v_{c}=20$~m/s.     

\begin{figure}[tb]
	\centering
	\includegraphics[width=\columnwidth]{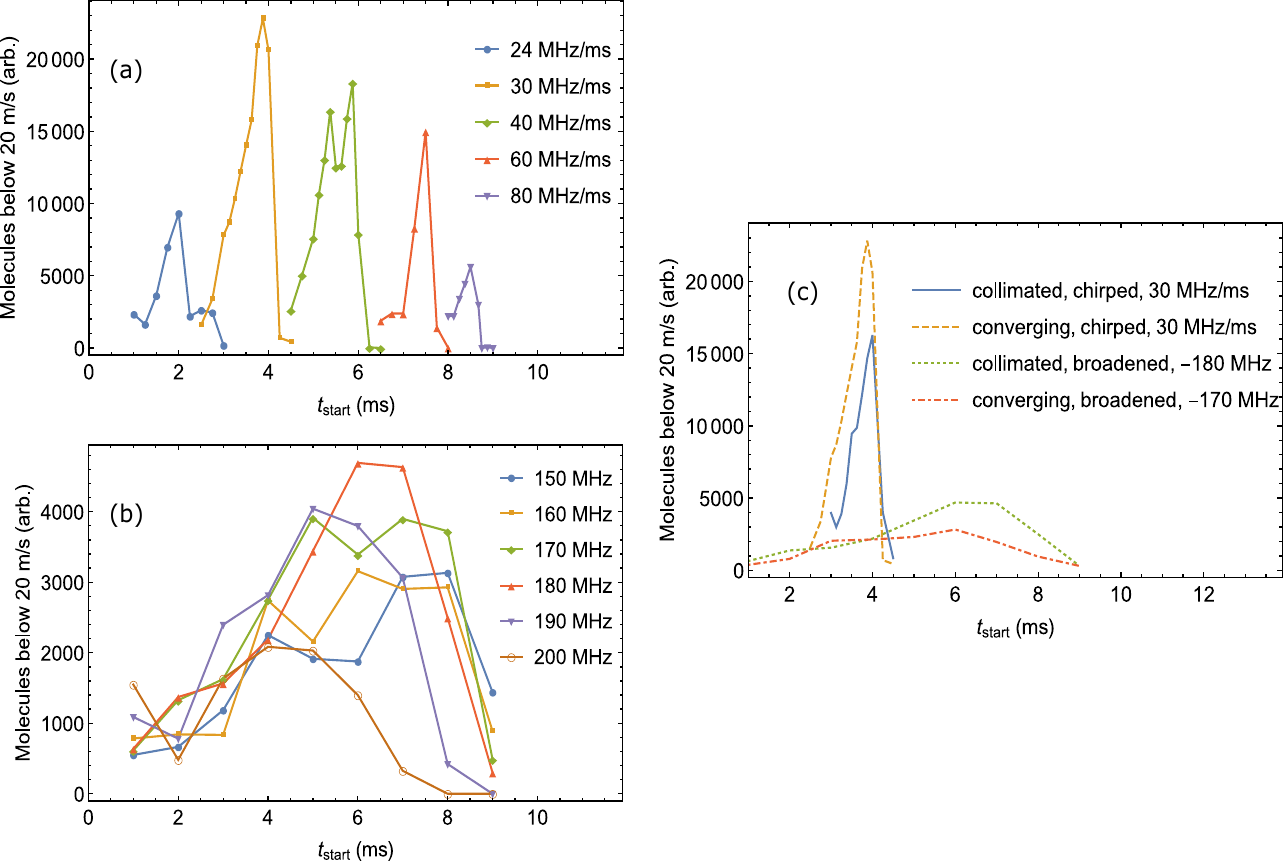}
	\caption{Result from simulations optimizing the number of molecules arriving at the MOT location below the expected capture velocity of 20~m/s. Here, the vertical scales are arbitrary. In all cases the slowing light is turned off at $t_{\rm{end}}=12$~ms and the chosen deceleration method is optimized by varying $t_{\rm{start}}$. (a) Chirped laser slowing using a converging cooling beam and an initial detuning of -335~MHz ($v_{\rm{start}}=178$~m/s) and various chirp rates.  (b) Frequency-broadened laser slowing using a collimated cooling laser and various overall detunings. (c) Comparison of the best parameter settings for the four cases of chirped, frequency-broadened, collimated-beam, and converging-beam laser slowing.}
	\label{Optimization_Figure}
\end{figure}

Figure~\ref{Optimization_Figure} shows the results of simulations aimed at optimising the number of molecules satisfying this MOT-loading metric. The five curves in Figure~\ref{Optimization_Figure}(a) compare chirped-frequency slowing using a converging beam with various values of $t_{\rm{start}}$ and $\beta$.  The best result is obtained with $\beta = 30$~MHz/ms and $t_{\rm{start}}=3.5$~ms.  The results are very sensitive to $t_{\rm{start}}$, as might be expected.  If the slowing light is turned on too late then, for a fixed chirp rate, no molecules are decelerated below the capture velocity.  If the slowing light is turned on too early, molecules decelerate too much and either diverge or are turned around before they reach the MOT location.  Figure~\ref{Optimization_Figure}(b) shows the optimization results using a collimated frequency-broadened slowing laser.  The results are a much weaker function of $t_{\rm{start}}$ than in Fig.~\ref{Optimization_Figure}(a) and are optimized at slightly later turn-on times.  The best result is obtained using $f_{\rm{offset}}= 180$~MHz and $t_{\rm{start}} = 6$~ms. Figure ~\ref{Optimization_Figure}(c) compares the best results of these optimization procedures for four cases: collimated-chirped, converging-chirped, collimated-broadened, and converging-broadened.  After this optimization, it is clear that chirped slowing outperforms frequency-broadened slowing in producing molecules at the MOT location and below the expected capture velocity.  Furthermore, this conclusion becomes even stronger if the MOT capture velocity is reduced. The optimized chirp method gives 4.5 times more molecules below $v_{c}$ than the optimized broadening method when $v_{c}=$20~m/s, and $>$20 times more when $v_{c}$=5~m/s. 

\section{Conclusions}

We have shown that a beam of CaF molecules can be slowed down either using the frequency-chirped method or the frequency-broadened method. By driving the B-X transition, which has exceptionally favourable branching ratios, the deceleration is rapid and efficient, requiring only two laser wavelengths, each with rf sidebands. Our study of losses to unaddressed states shows that $\sim 3\times 10^{4}$ photons per molecule can be scattered before half are lost from the cooling cycle. Molecules scattering this many photons would be slowed by 390~m/s, which is far greater than needed to bring molecules to rest from a typical buffer-gas-cooled source. For both slowing methods the dominant loss mechanism is the increased divergence of the slowed molecules. Hence, it is best to minimize the distance that the molecules have to travel at low speed, and so they should reach their final velocity as late as possible, i.e. when they reach the detector or the MOT volume. The frequency-broadened method is not good at achieving this because all molecules start slowing as soon as the light is turned on, and many reach low velocity too early. The chirped method is more efficient because the the slower molecules join the slowing process later on. For this reason, while the two methods produce a similar number of slow molecules down to about 50~m/s, the chirped method gives far more molecules at lower speeds, e.g. about ten times more at 20~m/s. This advantage is especially important for loading a MOT where the capture velocity is likely to be 20~m/s or less. We find that the chirped method yields more slow molecules when the slowing light converges towards the molecular source, especially for the lower velocities. Using this method, we produce approximately $10^{6}$ molecules with speeds in the narrow range 15$\pm$2.5~m/s. Thus, our method appears very well suited for loading a MOT. The chirped method also greatly compresses the velocity distribution, and it provides very precise velocity control. When combined with a short region of transverse laser cooling \cite{Shuman2010} near the source, our method will produce an intense, collimated, slow and velocity-controlled beam that could improve the precision of measurements that test fundamental physics.

\ack
The research leading to these results has received funding from EPSRC under grants EP/I012044 and EP/M027716, and from the European Research Council under the European Union's Seventh Framework Programme (FP7/2007-2013) / ERC grant agreement 320789.

\appendix

\section{Accuracy of the method for determining velocity distributions}\label{app}

Our method for determining velocity distributions is described in Sec.~\ref{analysisSection}. In this Appendix, we discuss in detail the accuracy of this method. The method must work perfectly if there is a unique correspondence between arrival time and velocity so that it is valid to assign all molecules arriving in any small time window to the mean velocity measured in that time window. However, molecules with different velocities may arrive at the same time if their journeys from source to detector differ in some way, so we wish to analyse the effect of that. We distinguish two ways that this can happen. The first is that molecules exit the source over a range of times. The second is that the force that acts may depend on other parameters such as the transverse position or transverse velocity of the molecule when it leaves the source. 

We consider first the case where molecules leave the source over a range of times. Let us define the exit time from the source, $t_0$, the transit time from source to detector, $\tau$, and the arrival time $t=\tau + t_{0}$. For now, we let the laser parameters be independent of time, so that a given initial velocity $u$ results in a specific final velocity $v$ and flight time $\tau$. Let these be related by $v = f(\tau)$ and the inverse, $\tau = g(v)$. The probability density function for a variable $x$ is $P_{x}(x)$. The time-of-flight profile measured 1.3~m from the source is $P_{t}(t)$ and the one measured 2.5~cm from the source is a good approximation to $P_{t_0}(t_0)$. 

The time-of-flight profile is
\begin{equation}
P_t(t) = \int P_\tau(t-t_0) P_{t_0}(t_0) d t_0 = (P_{t_0} \ast P_\tau)(t),
\end{equation}
where $\ast$ is the convolution operator. Thus, the distribution of transit times, $P_\tau(\tau)$, can be obtained from the data by the deconvolution of $P_t$ with $P_{t_0}$. The velocity distribution is related to $P_\tau(\tau)$ through a change of variables:
\begin{equation}
P_v(v) = P_\tau(g(v))\left|\frac{d g}{d v}\right|.\label{Pv}
\end{equation}
We do not measure $g(v)$ directly. Instead, we measure the mean velocity of molecules that arrive in a small time window centred at $t$, $\bar{v} = p(t)$. This can be expressed as
\begin{equation}
p(t) = \frac{\int f(t-t_0)P_\tau(t-t_0)P_{t_0}(t_0) d t_0}{\int P_\tau(t-t_0)P_{t_0}(t_0) d t_0}.
\end{equation}
Thus, we can write
\begin{equation}
p(t) P_t(t) = (P_{t_0} \ast f P_{\tau})(t).
\end{equation}

We now have the algorithm for determining the velocity distribution from the measured data: (i) Calculate $P_\tau$ by a deconvolution of $P_t$ with $P_{t_0}$; (ii) Calculate $f(t)$ by taking a deconvolution of the product $p P_t$ with $P_{t_0}$, and then dividing by $P_\tau$; (iii) Invert $f(t)$ to obtain $g(v)$; (iv) Take the derivative of $g(v)$; (v) Use Eq.(\ref{Pv}).

In our experiment, $P_{t_0}$ has a very narrow width - the distribution we measure at 2.5~cm has a FWHM of 280~$\mu$s, and the distribution at the source must be even narrower. Using the measured velocity distribution of the unslowed beam, we infer a FWHM at the source of 240~$\mu$s. This width is very small compared to any of the times $t$ where $P_t(t)$ is significant, and is also very small compared to the widths of any features in $P_t(t)$. As a result, the deconvolution steps have a negligible effect. In this limit,
\begin{equation}
P_v(v) \approx P_t(q(v))\left|\frac{d q}{d v}\right|,\label{PvApprox}
\end{equation}
where $t = q(\bar{v})$ is the inverse function to $p(t)$, and the approximation is exact in the limit that $P_{t_0}(t_0) = \delta(t_0)$. This is the result we use for all our data and, as we shall see below, it is very accurate for our experiment.

Our source emits a narrower temporal distribution than is typical of most buffer gas sources. To evaluate the accuracy of our analysis method when the source emits a longer pulse, we test it on synthetic data. To generate this data, we first create molecules at the source with initial velocities drawn at random from a normal distribution whose mean and width are equal to those we measure in the experiment, and with exit times drawn from a normal distribution with zero mean and FWHM $\Delta t$.  The molecules are then subject to an acceleration function $a = a_0/(1+(v-v_0)^{2}/w^{2})$, where we choose $a_0 = -10^{4}$~m/s$^{2}$, $v_{0} = 80$~m/s and $w = 10$~m/s. We solve the equation of motion for each molecule to generate the exact arrival time and velocity distributions in a plane 1~m from the source. We also determine the mean velocity in a set of time windows, just as in the experiment. We then apply the same analysis routine to the synthetic data as used for the real data, and compare the velocity distribution determined this way to the exact distribution.

\begin{figure}[tb]
	\centering
	\includegraphics[width=0.8\columnwidth]{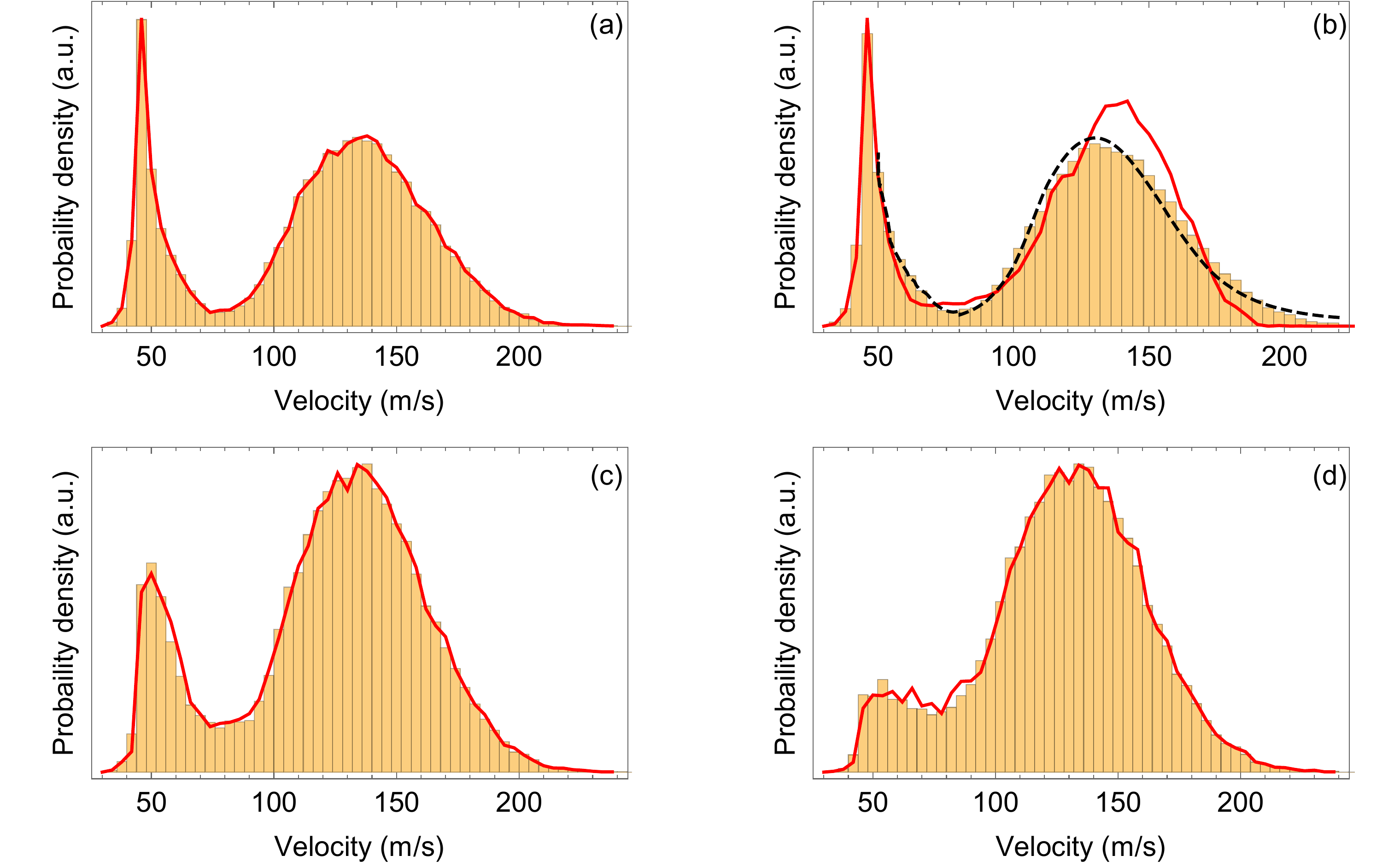}
	\caption{\label{MethodTests} Accuracy of using Eq.~(\ref{PvApprox}) to determine velocity distributions, assessed using model data. Histograms: exact distributions. Red lines: results using Eq.~(\ref{PvApprox}). Black dashed line in (b): distribution obtained using Eq.~(\ref{Pv}), the full analysis method. The parameters used in each case are described in the text. }
\end{figure}

Figure \ref{MethodTests}(a) shows this comparison in the case where we set $\Delta t = 240$~$\mu$s, as in the experiment. The histogram is the exact velocity distribution, and the line shows the distribution from Eq.~(\ref{PvApprox}). As expected from the argument above, there is no noticeable difference between the two. The largest difference in any velocity bin is 1.9\% of the amplitude of the undecelerated distribution, and the deviations in most bins are much smaller than this. Figure \ref{MethodTests}(b) shows the same comparison in the case where $\Delta t$ is 10 times larger. In this case, the distribution from Eq.~(\ref{PvApprox}) deviates considerably from the true one, especially for high velocities. This is to be expected since the arrival time is comparable to $\Delta t$ for these faster molecules. Interestingly, the analysis method still works well for the narrow distribution of slowed molecules which are the ones of most interest. This is because these molecules take a long time to reach the detector, and because the narrow peak in the velocity distribution does not correspond to any narrow features in the time-of-flight profile. On the contrary, the sharp feature in the velocity distribution arises because molecules arriving over a wide range of times all have very similiar velocities. The result of applying the full algorithm described above is shown by the dashed line in Fig.~\ref{MethodTests}(b) and does indeed give a better approximation to the true distribution in this case where the range of exit times is broad. We note that deconvolution algorithms often generate artificial oscillations in the result, especially where there are sudden changes in gradient, and that the analysis algorithm can become unstable when that occurs. We find that this happens at the low velocities where the sharp peak occurs, and so we only plot the result over the range where the algorithm is stable. Fortunately, the algorithm works well over the whole velocity range where the approximate method is inaccurate.

We have also compared the exact velocity distribution with the one determined from Eq.~(\ref{PvApprox}) for the case where the acceleration function is time-dependent. For this comparison, the acceleration acts only for times between 1 and 7~ms, and the resonant velocity $v_{0}$ is chirped downwards in time from 180 to 60~m/s, similar to the experiment. We use the narrow temporal source distribution of the experiment. Once again, we find that our analysis method reproduces the correct velocity distribution to very high accuracy.

We turn now to the possibility that molecules arriving in a small time window may have a spread of velocities because the integrated force depends on a parameter that differs between molecules, such as the transverse position or transverse velocity at the source. We use again our numerical model of the analysis method to examine the effect of this. We consider the case where molecules have a range of transverse positions $x$, but no transverse velocity. We modify the acceleration function so that it drops off with transverse displacement: $a = a_{0} \exp(-x^{2})/(1+(v-v_0)^{2}/w^{2})$. We produce the initial set of molecules as before, with $\Delta t = 240$~$\mu$s, and draw the dimensionless transverse displacement $x$ at random from a normal distribution with a full width at half maximum of 2. This samples a wider range of decelerations than the molecules experience in the experiment. There, the molecules that we detect travel close to the centre of the laser beam, where the intensity is high and the force is strongly saturated. Figure \ref{MethodTests}(c) shows the result for this case. We see that the range of forces broadens the peak of slow molecules, and that the analysis method accurately recovers the correct velocity distribution. Figure \ref{MethodTests}(d) shows the result when we choose instead an initial distribution which is uniform in the range $0<x<3$. This broadens and flattens the slow peak even further, and our analysis method still recovers the correct distribution. We have experimented with a range of different models for how the force and the initial distribution might vary, always finding that the analysis method is accurate.

\section*{References}
\bibliography{bib}

\begin{thebibliography}{10}

\bibitem{vandeMeerakker2012}
S.Y.T. van~de Meerakker, H.L. Bethlem, N.~Vanhaecke, and G.~Meijer.
\newblock Manipulation and control of molecular beams.
\newblock {\em Chem. Rev.}, 112:4828, 2012.

\bibitem{Hudson2011}
J.~J. Hudson, D.~M. Kara, I.~J. Smallman, B.~E. Sauer, M.~R. Tarbutt, and E.~A.
  Hinds.
\newblock Improved measurement of the shape of the electron.
\newblock {\em Nature}, 473:493, 2011.

\bibitem{Baron2014}
J.~Baron, W.~C. Campbell, D.~DeMille, J.~M. Doyle, G.~Gabrielse, Y.~V.
  Gurevich, P.~W. Hess, N.~R. Hutzler, E.~Kirilov, I.~Kozyryev, B.~R. O'Leary,
  C.~D. Panda, M.~F. Parsons, E.~S. Petrik, B.~Spaun, A.~C. Vutha, and A.~D.
  West.
\newblock Order of magnitude smaller limit on the electric dipole moment of the
  electron.
\newblock {\em Science}, 343(6168):269--272, 2014.

\bibitem{Cahn2014}
S.~B. Cahn, J.~Ammon, E.~Kirilov, Y.~V. Gurevich, D.~Murphree, R.~Paolino,
  D.~A. Rahmlow, M.~G. Kozlov, and D.~DeMille.
\newblock Zeeman-tuned rotational level-crossing spectroscopy in a diatomic
  free radical.
\newblock {\em Phys. Rev. Lett.}, 112:163002, 2014.

\bibitem{Daussy1999}
Ch. Daussy, T.~Marrel, A.~Amy-Klein, C.~T. Nguyen, Ch.~J. Bord\'e, and Ch.
  Chardonnet.
\newblock Limit on the parity nonconserving energy difference between the
  enantiomers of a chiral molecule by laser spectroscopy.
\newblock {\em Phys. Rev. Lett.}, 83:1554, 1999.

\bibitem{Tokunaga2013}
S.K. Tokunaga, C.~Stoeffler, F.~Auguste, A.~Shelkovnikov, C.~Daussy,
  A.~Amy-Klein, C.~Chardonnet, and B.~Darqui\'e.
\newblock Probing weak force-induced parity violation by high-resolution
  mid-infrared molecular spectroscopy.
\newblock {\em Mol. Phys.}, 111:2363, 2013.

\bibitem{Shelkovnikov2008}
A.~Shelkovnikov, R.~J. Butcher, C.~Chardonnet, and A.~Amy-Klein.
\newblock Stability of the proton-to-electron mass ratio.
\newblock {\em Phys. Rev. Lett.}, 100:150801, 2008.

\bibitem{Hudson2006}
E.~R. Hudson, H.~J. Lewandowski, B.~C. Sawyer, and J.~Ye.
\newblock Cold molecule spectroscopy for constraining the evolution of the fine
  structure constant.
\newblock {\em Phys. Rev. Lett.}, 96:143004, 2006.

\bibitem{Truppe2013}
S.~Truppe, R.~J. Hendricks, S.~K. Tokunaga, H.~J. Lewandowski, M.~G. Kozlov,
  C.~Henkel, E.~A. Hinds, and M.~R. Tarbutt.
\newblock A search for varying fundamental constants using {Hertz}-level
  frequency measurements of cold {CH}.
\newblock {\em Nat. Commun.}, 4:2600, 2013.

\bibitem{Salumbides2011}
E.J. Salumbides, G.D. Dickenson, T.I. Ivanov, and W.~Ubachs.
\newblock {QED} effects in molecules: test on rotational quantum states in
  {H$_2$}.
\newblock {\em Phys. Rev. Lett.}, 107:043005, 2011.

\bibitem{Bethlem1999}
H.~L. Bethlem, G.~Berden, and G.~Meijer.
\newblock Decelerating neutral dipolar molecules.
\newblock {\em Phys. Rev. Lett.}, 83:1558, 1999.

\bibitem{Osterwalder2010}
A.~Osterwalder, S.~A. Meek, G.~Hammer, H.~Haak, and G.~Meijer.
\newblock Deceleration of neutral molecules in macroscopic traveling traps.
\newblock {\em Phys. Rev. A}, 81:051401(R), 2010.

\bibitem{Fulton2004}
R.~Fulton, A.~I. Bishop, and P.~F. Barker.
\newblock Optical {S}tark decelerator for molecules.
\newblock {\em Phys. Rev. Lett.}, 93:243004, 2004.

\bibitem{Narevicius2008}
E.~Narevicius, A.~Libson, C.~G. Parthey, I.~Chavez, J.~Narevicius, U.~Even, and
  M.~G. Raizen.
\newblock Stopping supersonic oxygen with a series of pulsed electromagnetic
  coils: a molecular coilgun.
\newblock {\em Phys. Rev. A}, 77:051401, 2008.

\bibitem{Chervenkov2014}
S.~Chervenkov, X.~Wu, J.~Bayerl, A.~Rohlfes, T.~Gantner, M.~Zeppenfeld, and
  G.~Rempe.
\newblock {Continuous centrifuge decelerator for polar molecules}.
\newblock {\em Phys. Rev. Lett.}, 112:013001, 2014.

\bibitem{Fitch2016}
N.~J. Fitch and M.~R. Tarbutt.
\newblock {Principles and design of a Zeeman-Sisyphus decelerator for molecular
  beams}.
\newblock {\em Chem. Phys. Chem.}, 17, 2016.

\bibitem{Perez2013}
M.~Quintero-Perez, P.~Jansen, T.~E. Wall, J.~E. van~den Berg, S.~Hoekstra, and
  H.~L. Bethlem.
\newblock Static trapping of polar molecules in a traveling wave decelerator.
\newblock {\em Phys. Rev. Lett.}, 110:133003, 2013.

\bibitem{Stuhl2012}
B.~K. Stuhl, M.~T. Hummon, M.~Yeo, G.~Quemener, and J.~L. Bohn.
\newblock Evaporative cooling of the dipolar hydroxyl radical.
\newblock {\em Nature}, 492:396--400, 2012.

\bibitem{Zeppenfeld2009}
M.~Zeppenfeld, M.~Motsch, P.~W.~H. Pinkse, and G.~Rempe.
\newblock {Optoelectrical cooling of polar molecules}.
\newblock {\em Phys. Rev. A}, 80:041401, 2009.

\bibitem{Zeppenfeld2012}
M.~Zeppenfeld, B.~G.~U. Englert, R.~Gl\"ockner, A.~Prehn, M.~Mielenz,
  C.~Sommer, L.~D. van Buuren, M.~Motsch, and G.~Rempe.
\newblock {Sisyphus cooling of electrically trapped polyatomic molecules}.
\newblock {\em Nature}, 491:570, 2012.

\bibitem{Prehn2016}
A.~Prehn, M.~Ibr\"ugger, R.~Gl\"ockner, G.~Rempe, and M.~Zeppenfeld.
\newblock {Optoelectrical cooling of polar molecules to submillikelvin
  temperatures}.
\newblock {\em Phys. Rev. Lett.}, 116:063005, 2016.

\bibitem{Tokunaga2011}
S.~K. Tokunaga, W.~Skomorowski, P.~S. Zuchowski, R.~Moszynski, J.~M. Hutson,
  E.~A. Hinds, and M.~R. Tarbutt.
\newblock Prospects for sympathetic cooling of molecules in electrostatic, ac,
  and microwave traps.
\newblock {\em Eur. Phys. J. D}, 65:141, 2011.

\bibitem{Lim2015}
J.~Lim, M.~D. Frye, J.~M. Hutson, and M.~R. Tarbutt.
\newblock Modeling sympathetic cooling of molecules by ultracold atoms.
\newblock {\em Phys. Rev. A}, 92:053419, 2015.

\bibitem{Shuman2010}
E.~S. Shuman, J.~F. Barry, and D.~DeMille.
\newblock {Laser cooling of a diatomic molecule.}
\newblock {\em Nature}, 467(7317):820--3, 2010.

\bibitem{Hummon2013}
Matthew~T. Hummon, Mark Yeo, Benjamin~K. Stuhl, Alejandra~L. Collopy, Yong Xia,
  and Jun Ye.
\newblock {2D Magneto-Optical Trapping of Diatomic Molecules}.
\newblock {\em Phys. Rev. Lett.}, 110(14):143001, 2013.

\bibitem{Barry2014}
J.~F. Barry, D.~J. McCarron, E.~B. Norrgard, M.~H. Steinecker, and D.~DeMille.
\newblock Magneto-optical trapping of a diatomic molecule.
\newblock {\em Nature}, 512:286--289, 2014.

\bibitem{McCarron2015}
D.~J. McCarron, E.~B. Norrgard, M.~H. Steinecker, and D.~DeMille.
\newblock Improved magneto-optical trapping of a diatomic molecule.
\newblock {\em New J. Phys.}, 17:035014, 2015.

\bibitem{Norrgard2016}
E.~B. Norrgard, D.~J. McCarron, M.~H. Steinecker, M.~R. Tarbutt, and
  D.~DeMille.
\newblock Submillikelvin dipolar molecules in a radio-frequency magneto-optical
  trap.
\newblock {\em Phys. Rev. Lett.}, 116:063004, 2016.

\bibitem{Tarbutt2015}
M.~R. Tarbutt and T.~C. Steimle.
\newblock {Modeling magneto-optical trapping of CaF molecules}.
\newblock {\em Phys. Rev. A}, 92:053401, 2015.

\bibitem{Barry2012}
J.~F. Barry, E.~S. Shuman, E.~B. Norrgard, and D.~DeMille.
\newblock {Laser Radiation Pressure Slowing of a Molecular Beam}.
\newblock {\em Phys. Rev. Lett.}, 108(10):103002, 2012.

\bibitem{Yeo2015}
Mark Yeo, Matthew~T Hummon, Alejandra~L Collopy, Bo~Yan, Boerge Hemmerling,
  Eunmi Chae, John~M Doyle, and Jun Ye.
\newblock {Rotational State Microwave Mixing for Laser Cooling of Complex
  Diatomic Molecules.}
\newblock {\em Phys. Rev. Lett.}, 114(22):223003, 2015.

\bibitem{Hemmerling2016}
Boerge Hemmerling, Eunmi Chae, Aakash Ravi, Loic Anderegg, Garrett~K Drayna,
  Nicholas~R Hutzler, Alejandra~L Collopy, Jun Ye, Wolfgang Ketterle, and
  John~M Doyle.
\newblock {Laser slowing of CaF molecules to near the capture velocity of a
  molecular MOT}.
\newblock {\em J. Phys. B}, page 174001, 2016.

\bibitem{Pelegrini2005}
Marina Pelegrini, Ciro~S. Vivacqua, Orlando Roberto-Neto, Fernando~R. Ornellas,
  and Francisco B.~C. Machado.
\newblock {Radiative transition probabilities and lifetimes for the band
  systems A$^{2}\Pi$ - X$^{2}\Sigma^{+}$ of the isovalent molecules BeF, MgF
  and CaF}.
\newblock {\em Braz. J. Phys.}, 35(4a):950--956, 2005.

\bibitem{Dulick1980}
Michael Dulick, Peter~F. Bernath, and Robert~W. Field.
\newblock {Rotational and vibrational analysis of the CaF B$^{2}\Sigma^{+}$ –
  X $^{2}\Sigma^{+}$ system}.
\newblock {\em Can. J. Phys.}, 58(5):703--712, 1980.

\bibitem{Childs1981}
W.J. Childs, G.~L. Goodman, and L.~S. Goodman.
\newblock {Precise determination of the $v$ and $N$ dependence of the
  spin-rotation and hyperfine interactions in the CaF X$^{2}\Sigma_{1/2}$
  ground state}.
\newblock {\em J. Mol. Spectrosc.}, 86:365, 1981.

\bibitem{Dagdigian1974}
P.~J. Dagdigian, H.~W. Cruse, and R.~N. Zare.
\newblock {Radiative lifetimes of the alkaline earth monohalides}.
\newblock {\em J. Chem. Phys.}, 60:2330, 1974.

\bibitem{Tarbutt2013b}
M.~R. Tarbutt, B.~E. Sauer, J.~J. Hudson, and E.~A. Hinds.
\newblock {Design for a fountain of YbF molecules to measure the electron's
  electric dipole moment}.
\newblock {\em New J. Phys.}, 15(5):053034, 2013.

\bibitem{Shuman2009}
E.~S. Shuman, J.~F. Barry, D.~R. Glenn, and D.~DeMille.
\newblock {Radiative Force from Optical Cycling on a Diatomic Molecule}.
\newblock {\em Phys. Rev. Lett.}, 103(22):223001, 2009.

\bibitem{Zhelyazkova2014}
V.~Zhelyazkova, A.~Cournol, T.~E. Wall, A.~Matsushima, J.~J. Hudson, E.~A.
  Hinds, M.~R. Tarbutt, and B.~E. Sauer.
\newblock {Laser cooling and slowing of CaF molecules}.
\newblock {\em Phys. Rev. A}, 89(5):053416, 2014.

\bibitem{Hutzler2012}
Nicholas~R Hutzler, Hsin-I Lu, and John~M Doyle.
\newblock {The buffer gas beam: an intense, cold, and slow source for atoms and
  molecules.}
\newblock {\em Chem. Rev.}, 112(9):4803--27, 2012.

\bibitem{Barry2011a}
J.~F. Barry, E.~S. Shuman, and D.~DeMille.
\newblock {A bright, slow cryogenic molecular beam source for free radicals}.
\newblock {\em Phys. Chem. Chem. Phys.}, 13(42):18936, 2011.

\bibitem{Bulleid2013}
N~E Bulleid, S~M Skoff, R~J Hendricks, B~E Sauer, E~A Hinds, and M~R Tarbutt.
\newblock {Characterization of a cryogenic beam source for atoms and
  molecules.}
\newblock {\em Phys. Chem. Chem. Phys.}, 15(29):12299--307, 2013.

\bibitem{Berkeland2002}
D.~J. Berkeland and M.~G. Boshier.
\newblock {Destabilization of dark states and optical spectroscopy in
  Zeeman-degenerate atomic systems}.
\newblock {\em Phys. Rev. A}, 65:033413, 2002.

\bibitem{Stuhl2008}
B.~K Stuhl, B.~C Sawyer, D.~Wang, and J.~Ye.
\newblock {Magneto-optical Trap for Polar Molecules}.
\newblock {\em Phys. Rev. Lett.}, 101(24):243002, 2008.

\bibitem{Tarbutt2015b}
M.~R. Tarbutt.
\newblock {Magneto-optical trapping forces for atoms and molecules with complex
  level structures}.
\newblock {\em New J. Phys.}, 17:015007, 2015.

\bibitem{Zhu1991}
M.~Zhu, C.~W. Oates, and J.~L. Hall.
\newblock {Continuous high-flux monovelocity atomic beam based on a broadband
  laser-cooling technique}.
\newblock {\em Phys. Rev. Lett.}, 67(1):46--49, 1991.

\bibitem{Kirste2012}
Moritz Kirste, Xingan Wang, Gerard Meijer, Koos~B Gubbels, Ad~van~der Avoird,
  Gerrit~C Groenenboom, and Sebastiaan Y~T van~de Meerakker.
\newblock {Communication: Magnetic dipole transitions in the OH A
  $^{2}\Sigma^{+}$+ - X $^{2}\Pi$ system}.
\newblock {\em J. Chem. Phys.}, 137(10):101102, 2012.

\end{thebibliography}

\end{document}